\documentclass[ALICE,manyauthors]{cernphprep}
\usepackage[comma,square,numbers,sort&compress]{natbib}
\usepackage{hyperref}
\usepackage{lineno}
\usepackage{xspace}
\usepackage{amssymb}
\usepackage{amsmath,bm}
\usepackage{mathrsfs}
\usepackage{epstopdf}
\DeclareGraphicsExtensions{.eps}
\usepackage{color}
\usepackage{xcolor}
\usepackage{graphicx}
\usepackage[T1]{fontenc}
\usepackage{orcidlink}

\newcommand{\tevt}{$\sqrt{s} =$ 13 TeV}

\newcommand{\pt}           {\ensuremath{p_{\rm T}}\xspace}

\newcommand{\nineH}        {$\sqrt{s}~=~0.9$~Te\kern-.1emV\xspace}
\newcommand{\seven}        {$\sqrt{s}~=~7$~Te\kern-.1emV\xspace}
\newcommand{\twoH}         {$\sqrt{s}~=~0.2$~Te\kern-.1emV\xspace}
\newcommand{\twosevensix}  {$\sqrt{s}~=~2.76$~Te\kern-.1emV\xspace}
\newcommand{\five}         {$\sqrt{s}~=~5.02$~Te\kern-.1emV\xspace}
\newcommand{\twosevensixnn}{$\sqrt{s_{\mathrm{NN}}}~=~2.76$~Te\kern-.1emV\xspace}
\newcommand{\fivenn}       {$\sqrt{s_{\mathrm{NN}}}~=~5.02$~Te\kern-.1emV\xspace}

\newcommand{\GeV}          {\text{Ge\kern-.1emV}\xspace}
\newcommand{\MeV}          {\text{Me\kern-.1emV}\xspace}
\newcommand {\tev}      {\text{Te\kern-.1emV}\xspace}

\newcommand {\meanpT}    {\ensuremath{\langle p_{\mathrm{T}} \kern-0.1em\rangle}\xspace}
\newcommand {\mean}[1]   {\ensuremath{\langle #1 \kern-0.1em\rangle}\xspace}

\newcommand {\MeanNpart} {\mbox{\ensuremath{< \kern-0.15em N_{part} \kern-0.15em >}}}

\newcommand {\mmom}     {\mbox{\rm MeV$\kern-0.15em /\kern-0.12em c$}}
\newcommand {\gmom}     {\mbox{\rm GeV$\kern-0.15em /\kern-0.12em c$}}
\newcommand {\mmass}    {\mbox{\rm MeV$\kern-0.15em /\kern-0.12em c^2$}}

\newcommand {\dg}       {\mbox{$\kern+0.1em ^\circ$}}

\newcommand{\rmLambdas}         {\ensuremath{\mathrm {\Lambda \kern-0.2em + \kern-0.2em \overline{\Lambda}}}\xspace}

\newcommand{\Omegac}{\ensuremath{\Omega_{\rm c}^{0}}\xspace}
\newcommand{\eOmega}{\ensuremath{{\rm e}\Omega }\xspace}

\begin{document}

%%%%%%%%%%%%%%%  Title page %%%%%%%%%%%%%%%%%%%%%%%%
\begin{titlepage}
% the dates below correspond to CERN approval
% please don't touch: EB chairs will take care
\PHyear{2024}       % required, will be obtained from CERN
\PHnumber{112}      % required, will be obtained from CERN
\PHdate{23 April}  % required, will be obtained from CERN
%%%%%%%%%%%%%%%%%%%%%%%%%%%%%%%%%%%%%%%%%%%%%%%%%%%%

%===========================================================================================

%%% Put your own title + short title here:

\title{Measurement of $\Omega^{0}_{\rm c}$ baryon production and branching-fraction ratio ${\rm BR}(\Omega^0_{\rm c} \rightarrow \Omega^{-}{\rm e}^{+}\nu_{\rm e})/ {\rm BR}(\Omega^0_{\rm c} \rightarrow \Omega^{-}{\pi}^{+})$ in pp collisions at $\sqrt{s} =$ 13~TeV}

\ShortTitle{Measurement of the $\Omega^{0}_{\rm c}$ branching-fraction ratio}   %  appears on left page headers

% 
%%% Do not change the next lines
\Collaboration{ALICE Collaboration\thanks{See Appendix~\ref{app:collab} for the list of collaboration members}}
\ShortAuthor{ALICE Collaboration} % appears on right page headers, do not change

\begin{abstract}

The inclusive production of the charm-strange baryon $\Omegac$ is measured for the first time via its semileptonic decay into $\Omega^{-}\rm e^{+}\nu_{e}$ at midrapidity ($|y|<0.8$) in proton--proton (pp) collisions at the centre-of-mass energy $\sqrt{s}=13$~TeV with the ALICE detector at the LHC. The transverse momentum ($p_{\rm T}$) differential cross section multiplied by the branching ratio is presented in the interval $2<p_{\rm T}<12~{\rm GeV}/c$. The branching-fraction ratio ${\rm BR}(\Omega^0_{\rm c} \rightarrow \Omega^{-}{\rm e}^{+}\nu_{\rm e})/ {\rm BR}(\Omega^0_{\rm c} \rightarrow \Omega^{-}{\pi}^{+})$ is measured to be 1.12 $\pm$ 0.22 (stat.) $\pm$ 0.27 (syst.). Comparisons with other experimental measurements, as well as with theoretical calculations, are presented.

\end{abstract}
\end{titlepage}

\setcounter{page}{2} %please do not remove this line

%%%%%%%%%%%%%%%%%%%%%%%%%%%%%%%%%%%%%%%%%%%%%%%%%%%%%%%%%%%%%%%%%%%%%%%%%%%%%%%%%%%%%
%%%%%%%%%%%%%%%%%%%%%%%%%          Begin the main text                  %%%%%%%%%%%%%%%%%%%%%%%%%
%%%%%%%%%%%%%%%%%%%%%%%%%%%%%%%%%%%%%%%%%%%%%%%%%%%%%%%%%%%%%%%%%%%%%%%%%%%%%%%%%%%%%

%%%======================================================================================
\section{Introduction}

Production measurements of heavy-flavour hadrons (i.e. hadrons containing charm or beauty quarks) in high-energy proton--proton (pp) collisions provide essential tests of calculations based on the quantum chromodynamics (QCD) factorisation approach~\cite{Cacciari:1998it}. These frameworks exploit the fact that the heavy-quark masses are much larger than the QCD energy scale, $\Lambda_{\rm QCD}$, to calculate the production of heavy-flavour hadrons as a convolution of three factors: i) the parton distribution functions (PDFs) of the incoming protons; ii) the cross section of the partonic hard scattering; and iii) the fragmentation functions that parametrise the non-perturbative evolution of a heavy quark into a given heavy-flavour hadron species. Heavy-quark hadronisation is typically studied via the measurement of hadron-to-hadron yield ratios,  because the PDFs and partonic scattering cross sections are common to the charm- or beauty- hadron species and, therefore, with appropriate choice of the scheme of calculation it can be cancelled out in the yield ratios.

At the LHC, extensive production measurements of  $\rm D^{0}$, $\rm D^{+}$, $\rm D^{+}_{s}$, $\rm D^{*+}$ charmed  mesons~\cite{ALICE:2019nxm,ALICE:2021mgk,ATLAS:2015igt,CMS:2017qjw,CMS:2018bwt,LHCb:2016ikn,LHCb:2015swx} and of $\Lambda_{\rm c}^{+}$, $\Sigma_{\rm c}^{0,++}$, $\Xi^{0,+}_{\rm c}$, $\Omega^{0}_{\rm c}$ charmed  baryons~\cite{ALICE:2020wfu,ALICE:2020wla,ALICE:2022exq,CMS:2019uws,ALICE:2021rzj,ALICE:2021bli,Xic05TeV,ALICE:OmegaC} have been conducted. The meson-to-meson production yield ratios, for both the respective prompt (i.e. produced in the hadronisation of charm quarks or from the decay of excited open charm and charmonium states) and non-prompt (coming from beauty hadron decays)  components of the D-meson production, are observed to be independent of the transverse momentum ($p_{\rm T}$) within uncertainties. These ratios are well described by perturbative calculations at next-to-leading order, with next-to-leading-log resummation~\cite{Kniehl:2005mk,Kniehl:2012ti,Cacciari:1998it,Cacciari:2012ny}, which incorporate fragmentation functions tuned on $\rm e^{+}e^{-}$ and $\rm e^{-}p$ collision measurements. In contrast, these calculations underestimate the observed enhancement of the production of baryons relative to mesons in hadronic collisions~\cite{ALICE:2021dhb,ALICE:2023sgl}, with respect to the same measurements performed in $\rm e^{+}e^{-}$ or $\rm e^{\pm}p$ collisions. Several models have been proposed to explain the baryon enhancement. They either include dynamical processes that are relevant in quark-and-gluon enriched systems (e.g. colour reconnection beyond leading colour approximation~\cite{Christiansen:2015yqa} and quark coalescence~\cite{Song:2018tpv,Minissale:2020bif}), or that treat hadronisation as a statistical process while considering a set of yet-unobserved higher-mass charm-baryon states~\cite{He:2019tik}. 

Currently, a significant limitation in interpreting the production results of heavier strange-charm baryons is the absence of precise branching ratio measurements. In the recent inclusive $\Omega^{0}_{\rm c} \rightarrow \Omega^{-}\pi^{+}$ measurement reported in Ref.~\cite{ALICE:OmegaC}, model calculations were multiplied by the branching ratio, \mbox{$\rm BR(\Omega_c^0\rightarrow\Omega^-\pi^+)$} $=(0.51^{+2.19}_{-0.31})\%$. This value was obtained by considering the estimate reported in Ref.~\cite{Hsiao:2020gtc} for the central value, and the envelope of the values (including their uncertainties) reported in Refs.~\cite{Hsiao:2020gtc, Gutsche:2018utw, Cheng:1996cs, Hu:2020nkg, Wang:2022zja} to determine the uncertainty. This large uncertainty limited the understanding of $\Omega^{0}_{\rm c}$ production, meaning it is, therefore,  imperative to measure its branching ratios. The Belle and CLEO Collaborations published measurements of ${\rm BR}(\Omega^0_{\rm c} \rightarrow \Omega^{-}{\rm e}^{+}\nu_{\rm e})/ {\rm BR}(\Omega^0_{\rm c} \rightarrow \Omega^{-}{\pi}^{+})$ and found \mbox{1.98 $\pm$ 0.13 (stat.) $\pm$ 0.08 (syst.)}~\cite{Belle:2021dgc} and 2.4 $\pm$ 1.1 (stat.) $\pm$ 0.2 (syst.)~\cite{CLEO:2002imi}, respectively. Model calculations based on the light-front approach and the light-cone sum rules predict lower values of the branching fraction ratio of $1.1 \pm 0.2$~\cite{Hsiao:2020gtc} and $0.71$~\cite{Aliev:2022gxi}, respectively. These models provide a way to relate the properties of hadrons, such as their masses, decay constants, and form factors, to fundamental QCD parameters and quark--gluon distributions within the hadrons. The hadronic part of the weak decay is parameterised in those models in terms of form factors, which belong to the non-perturbative region of QCD. Another tool to study the charmed baryon decays is based on the flavor symmetry of $\rm SU(3)_{f}$ in the quark model, which allows the calculations of decay modes and relative probabilities of charmed baryon decays~\cite{Geng:2019bfz,Geng:2018upx,Liu:2023dvg}. By applying the effective colour approach under the framework of $\rm SU(3)_{f}$ symmetry, a branching fraction ratio of 1.35~\cite{Liu:2023dvg} is computed. The differences between the model calculations and the experimental values underscore the need for further experimental measurements and theoretical developments.

This letter presents the first \pt-differential inclusive production measurement of the semileptonic decay channel $\Omega^0_{\rm c} \rightarrow \Omega^{-}{\rm e}^{+}\nu_{\rm e}$ and the branching-fraction ratio ${\rm BR}(\Omega^0_{\rm c} \rightarrow \Omega^{-}{\rm e}^{+}\nu_{\rm e})/ {\rm BR}(\Omega^0_{\rm c} \rightarrow \Omega^{-}{\pi}^{+})$ at midrapidity ($|y| < 0.8$) in pp collisions at the centre-of-mass energy \tevt. The $\Omega^0_{\rm c}$ baryon is reconstructed together with its charge conjugate in three \pt intervals, 2~$< \pt <$~4~GeV$/c$, 4~$< \pt <$~6~GeV$/c$, and 6~$< \pt <$~12~GeV$/c$.

%======================================================================================
\section{Experimental setup and data samples}

The ALICE experiment and its performance are presented in detail in Refs.~\cite{Alice_performance1, Alice_performance2}. The main detectors used in this analysis are the Inner Tracking System (ITS)~\cite{DetectorITS}, the Time Projection Chamber (TPC)~\cite{DetectorTPC}, and the Time-Of-Flight detector (TOF)~\cite{DetectorTOF} for vertexing, tracking, and particle identification (PID) purposes. They are located in the central barrel covering the pseudorapidity interval ($|\eta| < 0.9$) and lie inside a solenoidal magnet that provides a magnetic field $B =$ 0.5 T parallel to the beam direction. The analysed data sample consists of pp collisions at \tevt \ recorded with a minimum-bias (MB) trigger. The MB trigger requires a pair of coincident signals in two scintillator arrays (V0)~\cite{DetectorV0}, which are located on both sides of the nominal interaction point along the beam direction.  Further offline event selection was applied to remove the contamination from beam--gas collisions and other machine-related backgrounds. These criteria were based on the timing information of the two V0 arrays and a selection on the correlation between clusters and tracklets reconstructed in the two innermost layers of the ITS (Silicon Pixel Detector, SPD). Only events with a reconstructed vertex position within $\pm 10$ cm along the beam axis from the nominal interaction point were analysed, to maintain a uniform ITS acceptance in pseudorapidity. The primary-vertex position was defined using tracks reconstructed in the TPC and ITS detectors. 
Events with multiple reconstructed primary vertices, which amount to about 1\% of the total event sample, were rejected to reduce the contamination from the superposition of several collisions within the same colliding bunches (pile-up events). After the aforementioned selection criteria, the data sample corresponds to an integrated luminosity $\mathcal{L}_{\rm int}=$~(32.08 $\pm$ 0.51)~${\rm nb^{-1}}$~\cite{ALICE-PUBLIC-2021-005}.

\section{Analysis method}

The \Omegac candidates were built by pairing an electron or positron candidate track with an $\Omega$ cascade candidate using a Kalman-Filter (KF) vertexing algorithm~\cite{kfparticle}. Charge conjugate modes are included everywhere unless otherwise stated.

The $\Omega$ candidates were reconstructed via the decay chain $\Omega^{-} \rightarrow \Lambda \rm K^{-}$ (BR = ($67.8 \pm 0.7) \%$), followed by the decay $\Lambda \rightarrow {\rm p}\pi^{-}$ (BR = ($63.9 \pm 0.5)\%$)~\cite{pdg2022}, exploiting the characteristic decay topology as reported in Refs.~\cite{Acharya:2019kyh, ALICE:OmegaC}. Charged-particle tracks used in this analysis were required to be within the pseudorapidity interval $|\eta| <$ 0.8 and to have a number of crossed TPC pad rows larger than 70 out of a maximum of 159. Particle identification (PID) selection was based on the differences between the measured and expected response for a given particle species hypothesis, in units of the detector resolution ($n\sigma_{\rm det}$). For proton, pion, and kaon tracks, a selection on the measured specific energy loss d$E/$d$x$ in the TPC of $|n\sigma_{\rm TPC}| < 4$ was applied for the respective particle hypothesis. An additional PID selection of $|n\sigma_{\rm TOF}| < 5$ was applied for the kaon candidates when information from the TOF detector was available. Tracks without TOF hits were identified using only the TPC information.

Electron candidate tracks were selected by requiring to have a minimum of three (out of a maximum of six) hits in the ITS with two in the SPD layers~\cite{Electron_ITS1,Electron_ITS2}, at least 50 clusters in the TPC, a number of crossed TPC pad rows larger than 70, and $p_{\rm T} > $ 0.5 GeV$/c$. These requirements help suppressing the contribution from short tracks, which are unlikely to originate from the $\Omega_{\rm c}^{0}$ decay. The dominant source of electron background is photon conversions. They were rejected by requiring hits in the SPD layers, minimising the effective material budget. The electron candidate tracks were identified by using d$E/$d$x$ and time-of-flight information in the TPC and TOF detectors, respectively. Two selection criteria on the PID of electron candidates, $|n\sigma_{\rm TPC}^{\rm e}| < 4 $ and $|n\sigma_{\rm TOF}^{\rm e}| < 5 $, were required. The remaining electrons steaming from photon conversion and those originating from Dalitz decays of neutral mesons were further rejected with an invariant-mass technique~\cite{prefilter1,Prefilter2}. The electron candidates were paired with opposite-sign tracks from the same event passing loose identification criteria ($|n\sigma_{\rm TPC}^{\rm e}| < 5 $ without any TOF requirement) and were rejected if they formed at least one $\rm e^{+}e^{-}$ pairs with an invariant mass smaller than 50 MeV$/c^{2}$. 

The $\Omegac$ candidates were selected by requiring the cosine of the opening angle between the electron and the $\Omega$ candidate tracks to be greater than 0 for $ 2 < p_{\rm T} < 4$ GeV$/c$, 0.25 for $ 4 < p_{\rm T} < 6$ GeV$/c$ and 0.5 for \mbox{$ 6 < p_{\rm T} < 12$ GeV$/c$}. The \pt dependence of this selection was chosen by looking at its correlation with the \eOmega-pair mass distribution in data and Monte Carlo (MC) simulations, minimising the rejection of signal candidates in the data. 
 
After applying the selections described above, further separation of the signal and background was based on the Boosted Decision Tree (BDT) algorithm implemented in the XGBoost library~\cite{XGBoost,hipe4ml}. Independent BDT models were trained for each \pt interval with a sample of signal and background candidates as performed in Refs~\cite{ALICE:2021kfc,ALICE:2022exq,ALICE:2021mgk}. For the reconstructed signal, \eOmega pairs from $\Omegac$ decays were obtained from simulations with the PYTHIA 8.2 event generator~\cite{PYTHIA82}. Each PYTHIA event was required to contain a $\rm c\overline{c}$ or $\rm b\overline{b}$ quark pair and \Omegac baryons were forced to decay into the $\Omega^{-}\rm e^{+}\nu_{e}$ channel. The mean proper lifetime of \Omegac baryons in the simulation was set to 268 fs based on the latest LHCb measurement~\cite{PRL_LHCb}. The transport of simulated particles within the detector was performed with the GEANT 3 package~\cite{Brun:1994aa}. The conditions of all the ALICE detectors in terms of active channels, gain, noise level, and alignment, as well as the evolution of the detector configurations during the data-taking period, were taken into account in the simulations. A mixed-event (ME) technique was used to increase statistics in the background sample. The ME technique exploited randomised sub-samples of the full dataset, using the same filtering selections described above, generating \eOmega pairs with the opposite charge. The ME background was obtained by correlating $\Omega$ candidates in an event with electron candidate tracks from other events with similar multiplicity and primary-vertex position along the beam direction. Exploiting a background sample using the same-charge \eOmega pairs in the same event (SE) was also tested. The resulting background distributions were found to be consistent with each other, and the SE pairs were used to normalise the more statistically abundant ME background sample.

The BDT training variables included topological properties of the decays and PID variables. The training variables related to the PID information were obtained by combining the ones coming from the TPC and TOF, $n\sigma^{\rm K, e}_{\rm combined} = \frac{1}{\sqrt{2} } \sqrt{\big(n\sigma_{\rm TPC}^{\rm K, e}\big)^2 + \big(n\sigma_{\rm TOF}^{\rm K, e}\big)^2 }$, on the electron and kaon tracks, respectively. The training variables describing the $\Omega$ decay topology were the distance of the closest approach (DCA) of the charged decay particles, the pointing angle of the reconstructed $\Omega$ momentum to the primary vertex, the $\chi^{2}_{\rm topo}/{\rm NDF}$, which, in this analysis, characterises whether the momentum vector of the $\Omega$ candidate points back to the reconstructed primary vertex of the event,
and the $\chi^{2}_{\rm geo}/{\rm NDF}$, 
which is related to the geometrical intersection of the daughter-particle trajectories, taking their uncertainties into account. The training variables related to the $\Lambda$ were the DCA to the primary vertex, the radial distance of the $\Lambda$ decay vertex from the beam axis, and the DCA between the decay particles. The BDT model output is a single response variable related to the probability that the candidate is a signal.
The selection on the BDT output was tuned in each \pt interval to maximise the expected statistical
significance, which was estimated using:  the expected signal obtained from the \Omegac production cross section reported in Ref.~\cite{ALICE:OmegaC} multiplied by the BDT selection efficiency and the
expected background estimated from the normalised ME.  The resulting BDT output thresholds were 0.86, 0.84, and 0.81 for the three \pt intervals, respectively.

\begin{figure}[t!]
\centering
\includegraphics[width=.45\textwidth]{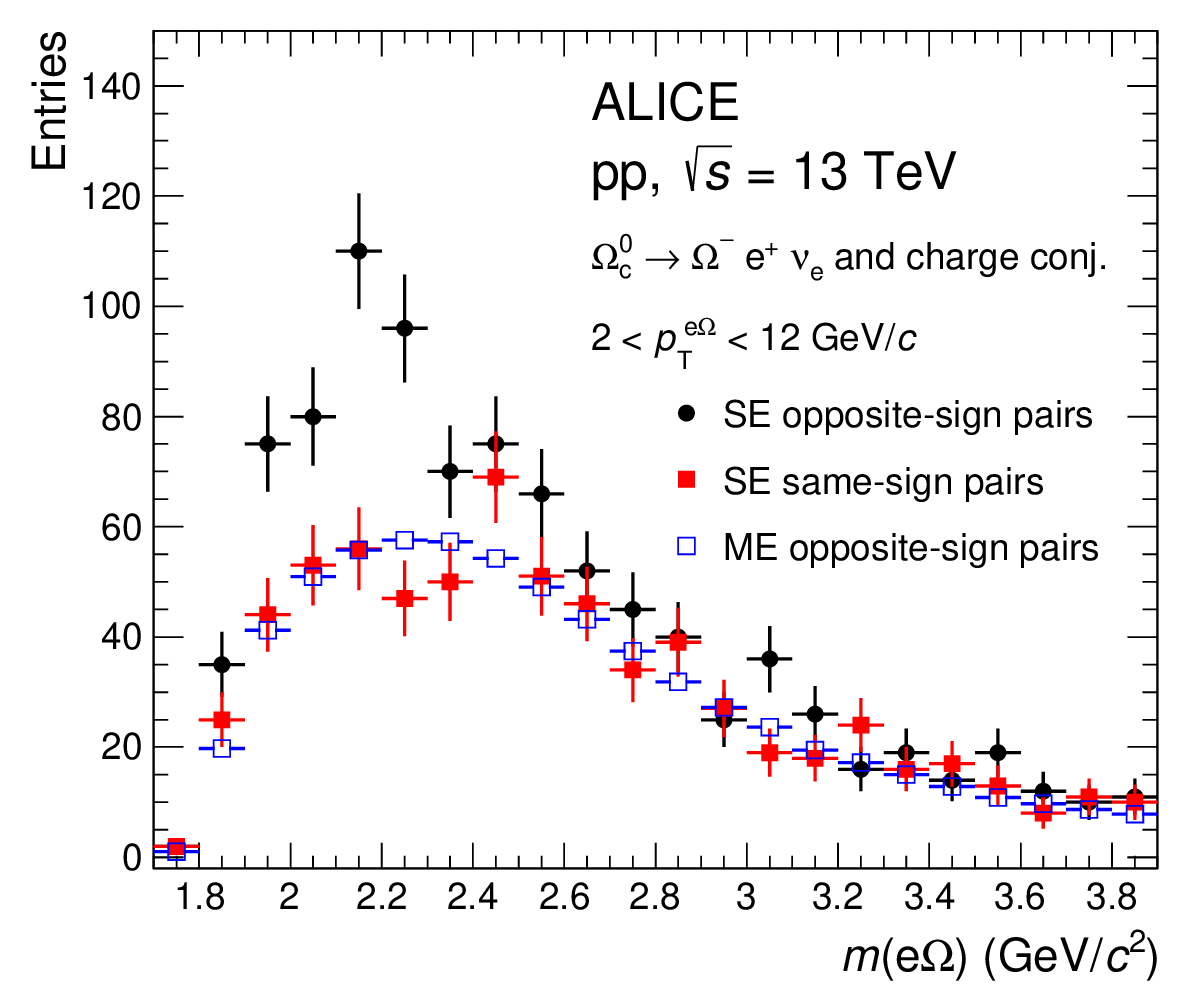}
\includegraphics[width=.45\textwidth]{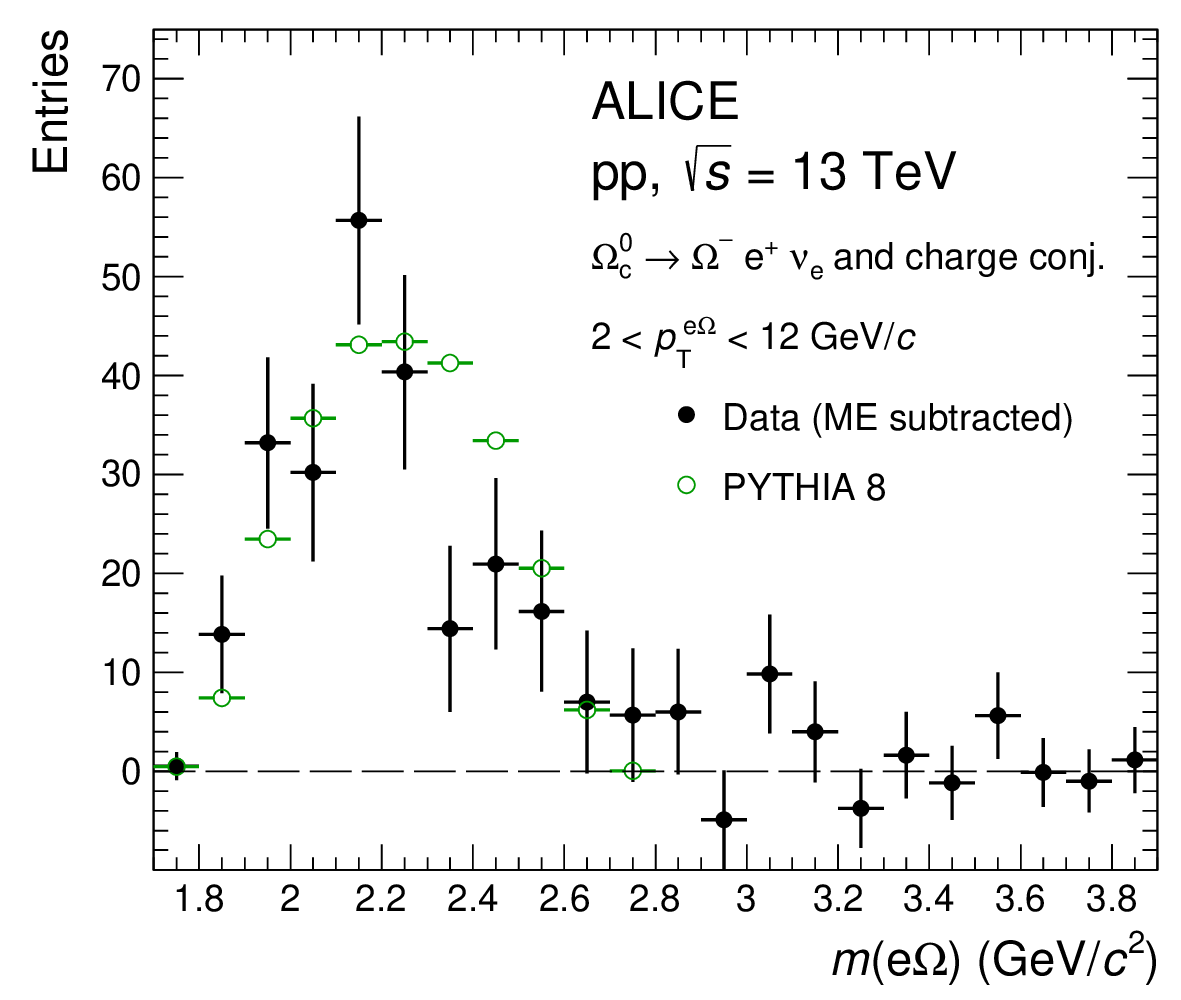}
\caption{Left panel: invariant-mass distribution of opposite-sign pairs (black solid circle marker) and same-sign pairs (red solid square marker) in SE, and opposite-sign pairs (blue open square marker) in ME. Right panel: invariant-mass distribution of the \Omegac candidates obtained by subtracting the opposite-sign charge \eOmega pairs in ME from the opposite-sign charge pairs in SE (black solid circle marker), and \eOmega opposite-sign charge pairs coming from \Omegac decay from PYTHIA 8 (green open circle marker).}
\label{fig1}
\end{figure}

\begin{figure}[t!]
\centering
\includegraphics[width=.45\textwidth]{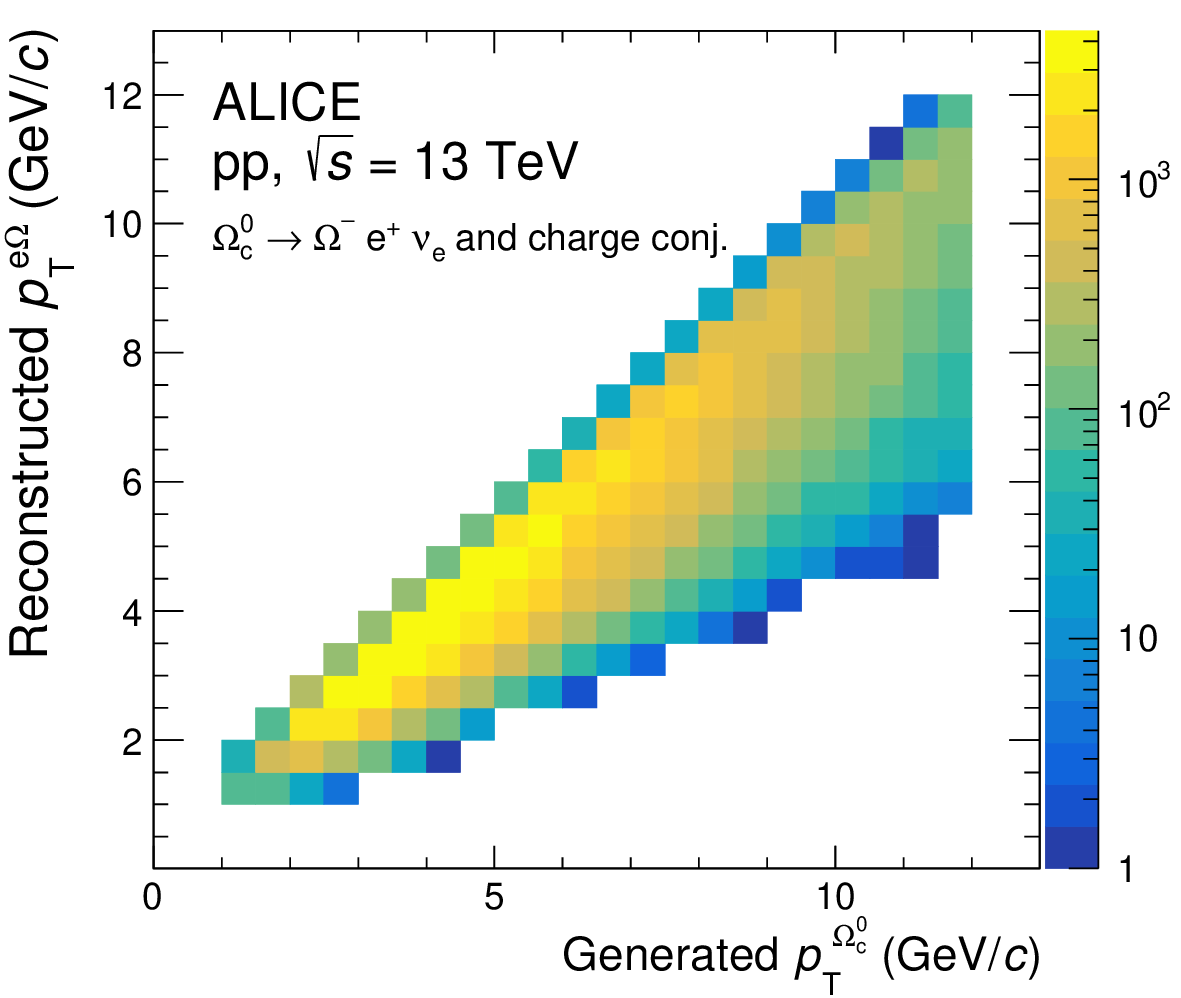}
\includegraphics[width=.45\textwidth]{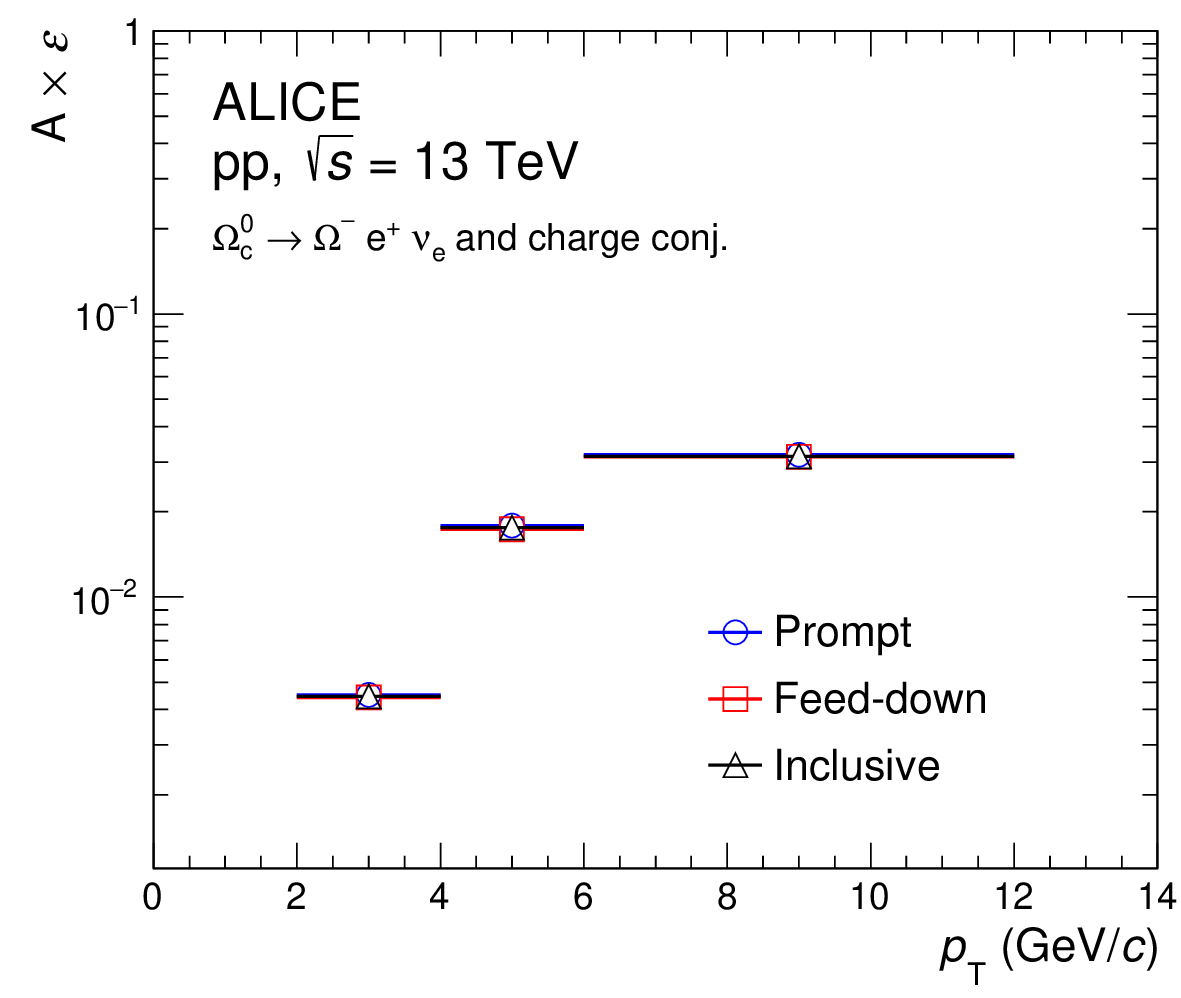}
\caption{Left panel: correlation matrix between the generated \Omegac  baryon $p_{\rm T}$ and the reconstructed opposite-sign charge pairs, obtained from the simulation based on PYTHIA 8 described in Ref.~\cite{Xic05TeV}. Right panel: product of $({\rm A}\times\varepsilon)$ for inclusive $\Omega^0_{\rm c}$ baryons in pp collisions at $\sqrt{s} =$ 13 TeV as a function of $p_{\rm T}$.}
\label{fig2}
\end{figure}

The left panel of Fig.~\ref{fig1} shows the invariant-mass distribution of \eOmega pairs in SE (same-sign and opposite-sign) and ME (opposite-sign) in the interval $2 < p_{\rm T}^{\rm e\Omega} < 12$ GeV$/c$. The raw yield was obtained by subtracting the combinatorial background computed using the ME technique from the invariant-mass distribution of \eOmega pairs with opposite-sign charge in the SE. The right panel of  Fig.~\ref{fig1} shows the invariant-mass distribution of \eOmega candidates, obtained after background subtraction, in comparison with \eOmega opposite-sign charge pairs coming from the \Omegac decay computed with the PYTHIA 8 event generator~\cite{PYTHIA82}. Only $\rm e\Omega$ pairs satisfying $1.7 < m_{\rm e\Omega} < 2.7$ GeV$/c^{2}$ were considered for further analysis. The number of reconstructed \eOmega signal pairs consists of 232~$\pm$~15 candidates. The missing momentum of the neutrino was corrected by using the Bayesian-unfolding technique~\cite{DAgostini:1994fjx} implemented in the RooUnfold package~\cite{Adye:2011gm}. The response matrix, which represents the correlation between the generated \Omegac and reconstructed \eOmega transverse momenta, used in the unfolding procedure is shown in the left panel of Fig.~\ref{fig2}. In this analysis, the Bayesian procedure requires two iterations to converge. Additional information on the unfolding procedure is explained in Ref.~\cite{Xic05TeV}. The response matrix was determined with the same simulation setup used for the BDT training.

The $p_{\rm T}$-differential production cross section of inclusive \Omegac \ baryons in the rapidity interval $|y|<0.8$ multiplied by the BR into the considered semileptonic decay channel was calculated from the yields obtained from the unfolding procedure as follows:

\begin{equation}
	{\rm BR}\times\frac{\text{d}^2\sigma^{\Omega^0_{\rm c}}}{\text{d}p_\mathrm{T}\text{d}y}= \frac{1}{2\Delta y\Delta p_\mathrm{T}}\times\frac{ N^{\Omega^0_{\rm c}}_\text{raw}}{({\rm A} \times\varepsilon)}\times\frac{1}{\mathcal{L}_{\rm int}},
	%\hspace{0.1cm},
	\label{eq::corrected_yield}
\end{equation}

where $N^{\Omega^0_{\rm c}}_{\rm raw}$ is the raw yield (sum of particles and antiparticles) in a given $p_{\rm T}$ and rapidity interval with width $\Delta p_\mathrm{T}$ and $\Delta y$. The factor $1/2$ takes into account that the raw yield includes both particles and antiparticles, while the cross section is given for particles only. The $\mathcal{L}_{\rm int}$ is the integrated luminosity. Since the feed-down contribution is not subtracted, the $({\rm A}\times\varepsilon)$ factor is the product of the acceptance and efficiency for inclusive $\Omega^0_{\rm c}$ baryons, where $\varepsilon$ accounts for the reconstruction and selection of the $\Omega^0_{\rm c}$ decay-product tracks and the $\Omega^0_{\rm c}$-candidate selection. The $({\rm A}\times\varepsilon)$ correction was obtained from a simulation with the same configuration as the one used for the BDT training and the response matrix. The $({\rm A}  \times\varepsilon)$ correction factors of prompt, beauty feed-down (non-prompt), and inclusive $\rm \Omega_c^0$ as a function of $p_{\rm T}$ are observed to be consistent with each other within uncertainties, because the selection variables used are not sensitive to the displacement by a few hundred micrometres of the prompt and beauty feed-down $\rm \Omega_c^0$ decay vertices from the collision point. The \Omegac-baryon $p_{\rm T}$ distribution from the PYTHIA 8 simulation was reweighted to match the true distribution, which was parametrised via a Tsallis fit to the differential production cross section of \Omegac as measured in Ref.~\cite{ALICE:OmegaC}. The right panel of Fig.~\ref{fig2} shows the product of the final $({\rm A}\times\varepsilon)$ correction factor for inclusive, prompt and feed-down \Omegac as a function of $p_{\rm T}$.

%%%======================================================================================
\section{Systematic uncertainties}

\begin{table}[tb!]
\caption{Contributions to the systematic uncertainty of the $\Omega^{0}_{\rm{c}}$ cross section for the \pt intervals \mbox{ $ 2 < p_{\rm T} < 4$ GeV$/c$}, $4 < p_{\rm T} < 6$ GeV$/c$, and $6 < p_{\rm T} < 12$ GeV$/c$. The global uncertainty on the luminosity is quoted separately and it is not added in quadrature to the other sources.}
	\begin{center}
	\renewcommand\arraystretch{1.4}
	\begin{tabular}{lcccc}
	\hline 
	\hline
	  \pt (GeV/$c$) & 2--4 & 4--6 & 6--12\\
	  \hline
      Raw-yield extraction &  10\% & 10\% &10\% \\
	  ITS--TPC matching efficiency &  2\% & 2\% & 2\%  \\
	  Track efficiency & 4\% & 4\% & 4\%  \\
	  Bayesian-unfolding iterations & 4\%&4\% & 4\% \\
	  Unfolding method & 4\% &4\% & 4\% \\
	  Response-matrix \pt range and binning	&  10\%  & 10\% & -- \\
	  BDT selection &  15\%&15\% & 15\% \\
	  Generated $p_{\rm T}$ shape & 10\% &2\% &1\% \\

	 \hline
     \hline
	 {Total systematic uncertainty }& 24\% & 22\% & 19\% \\
	\hline
      Luminosity & & 1.6\% & \\
	\hline
    \hline
	\end{tabular}
	\end{center}
	\label{tab:systotal}
\end{table}

The different contributions to the total systematic uncertainty of the \Omegac production cross section in \mbox{ $ 2 < p_{\rm T} < 12$~GeV$/c$} are summarised in Table~\ref{tab:systotal}. The various systematic sources were defined as the RMS of the distribution of the corrected yields obtained from the reported variations, if not differently specified. 

The systematic uncertainty on the raw-yield extraction was evaluated by investigating possible contaminations to \eOmega pairs with the opposite charge from other decays. The contamination of different decay channels mentioned in Refs.~\cite{CLEO:2002imi,Faller:2015oma} are the $\Omega_{\rm c}^{0} \rightarrow \Xi^{+}_{\rm c} \rm e^{-} \nu_{e}$, $\Xi^{0}_{\rm c} \rightarrow \Omega^{-}\rm e^{+}\nu_{e} K^{0} $,  $\Xi^{+}_{\rm c} \rightarrow \Omega^{-}\rm e^{+}\nu_{e} K^{+} $, and $\Omega_{\rm c}^{0} \rightarrow \Omega^{-}\rm e^{+}\nu_{e} \pi^{+} \pi^{-}$. From PYTHIA 8 simulation studies, it was found that these decays mainly contribute to a mass region below 2.2~GeV$/c^2$. A maximum variation of 10\% at the corrected yield level was found by varying the lower limit of the integration mass range for the signal extraction in the \eOmega mass from 1.7 to 2.2~GeV$/c^{2}$, which was assigned as systematic uncertainty. Note that those decay channels are not experimentally observed, therefore the Belle~\cite{Belle:2021dgc} and CLEO~\cite{CLEO:2002imi} Collaborations do not correct for it in addition to not assigning a corresponding systematic uncertainty.

The systematic uncertainty on the tracking efficiency was determined by comparing the matching efficiency of prolonging a track from the TPC to the ITS in data and simulation, and by varying the track quality selection criteria. The uncertainty on the matching efficiency, defined as the relative difference in the ITS--TPC matching efficiency between the data and simulation, affected only the electron track. For the tracks of the $\Omega$ decay particles, the prolongation to the ITS was not required. The uncertainties on electron tracks were propagated to the \Omegac candidates according to the decay kinematics, resulting in an uncertainty of 2\%.  The second contribution to the track reconstruction was estimated by varying the track quality selection criteria and 4\% uncertainty was assigned.

The systematic uncertainty on the unfolding procedure was determined by considering three contributions.  The first contribution was due to the regularisation procedure in the Bayesian unfolding. It was estimated by varying the iteration number between 2 and 5, and an uncertainty of 4\% was assigned. The second contribution was estimated by unfolding with the singular value decomposition algorithm~\cite{Hocker:1995kb}, and a 4\% uncertainty was assigned, independent of \Omegac $p_{\rm T}$. 
The third source was related to the sensitivity of the unfolding to bin edge effects and was estimated by varying the \pt range and the binning of the response matrix.
An uncertainty of 10\% was assigned in the interval $ 2 < p_{\rm T} < 6$ GeV$/c$. At higher \pt, no variations were observed when using finer \pt intervals in the unfolding procedure.

The systematic uncertainty on the selection efficiency originates from imperfections in the description of the detector response and alignment in the simulation. It was estimated from the ratios of the corrected yields obtained by varying the selections on the BDT outputs, which results in modification of the efficiencies, raw yield, and background values. The systematic evaluation was extended using a BDT model with different training variables (no PID included in the training) and preselection (PID selections were varied when not included in the BDT). A value of 15\% was assigned as systematic uncertainty.

The systematic uncertainty  due to the difference in the shape of the true and generated \Omegac $p_{\rm T}$ distributions was estimated by varying the Tsallis fit
used to determine the \pt weights within the statistical and \pt uncorrelated uncertainties. The assigned uncertainty, defined as the maximum variation observed, was 10\% in the interval $ 2 < p_{\rm T} < 4$ GeV$/c$, 2\% in the interval $ 4 < p_{\rm T} < 6$ GeV$/c$, and 1\% for the highest $p_{\rm T}$ interval.

All systematic uncertainties are considered uncorrelated and summed in quadrature to obtain the total systematic uncertainty. The production cross section has an additional global normalisation uncertainty of 1.6\% due to the uncertainties of the integrated luminosity~\cite{ALICE-PUBLIC-2021-005}. 

%%%======================================================================================
\section{Results}

\begin{figure}[t!]
\centering
\includegraphics[width=.55\textwidth]{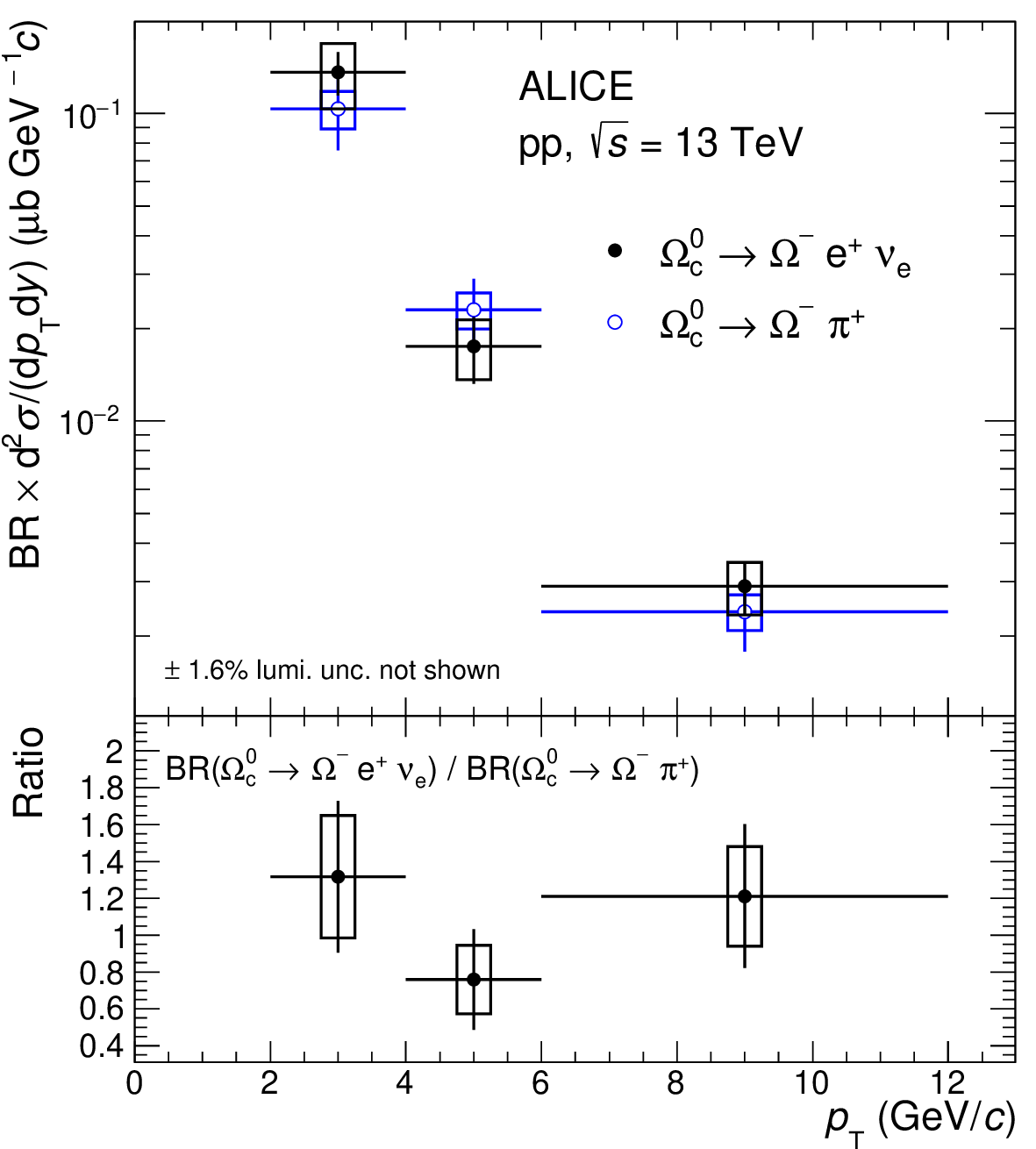}
\caption{Top panel: $p_{\rm T}$-differential production cross sections of inclusive \Omegac baryons multiplied by the branching ratios (BR) into $\Omega^{-}{\rm e}^{+}\nu_{\rm e}$ and $\Omega^{-}\pi^{+}$ ~\cite{ALICE:OmegaC} in pp collisions at \tevt. Bottom panel: $p_{\rm T}$-differential branching-fraction ratio ${\rm BR}(\Omega^0_{\rm c} \rightarrow \Omega^{-}{\rm e}^{+}\nu_{\rm e})/ {\rm BR}(\Omega^0_{\rm c} \rightarrow \Omega^{-}{\pi}^{+})$.}
\label{fig3}
\end{figure}

The $p_{\rm T}$-differential cross section of inclusive \Omegac baryon production multiplied by the branching ratio into $\Omega^{-}{\rm e}^{+}\nu_{\rm e}$, in pp collision at \tevt, measured in rapidity interval $|y| < $ 0.8 and the $p_{\rm T}$ interval $ 2 < p_{\rm T} < 12 $ GeV$/c$, is shown in the top panel of Fig.~\ref{fig3}.  It is compared with the previously published measurements of inclusive \Omegac baryon production in the hadronic decay channel $\Omegac \rightarrow \Omega^{-}\pi^{+}$. The error bars and boxes represent the statistical and systematic uncertainty, respectively. The uncertainty of the integrated luminosity is not included in the boxes. In the bottom panel of Fig.~\ref{fig3} the branching-fraction ratio ${\rm BR}(\Omega^0_{\rm c} \rightarrow \Omega^{-}{\rm e}^{+}\nu_{\rm e})/ {\rm BR}(\Omega^0_{\rm c} \rightarrow \Omega^{-}{\pi}^{+})$ is shown as function of $p_{\rm T}$.  The systematic uncertainties on the branching-fraction ratio were calculated assuming all the uncertainties between the two measurements as uncorrelated, except for the ITS--TPC matching efficiency, track quality selection, and the MC $p_{\rm T}$ shape. The uncertainty of the luminosity cancels in the ratio, as it is fully correlated.

\begin{figure}[t!]
\centering
\includegraphics[width=.55\textwidth]{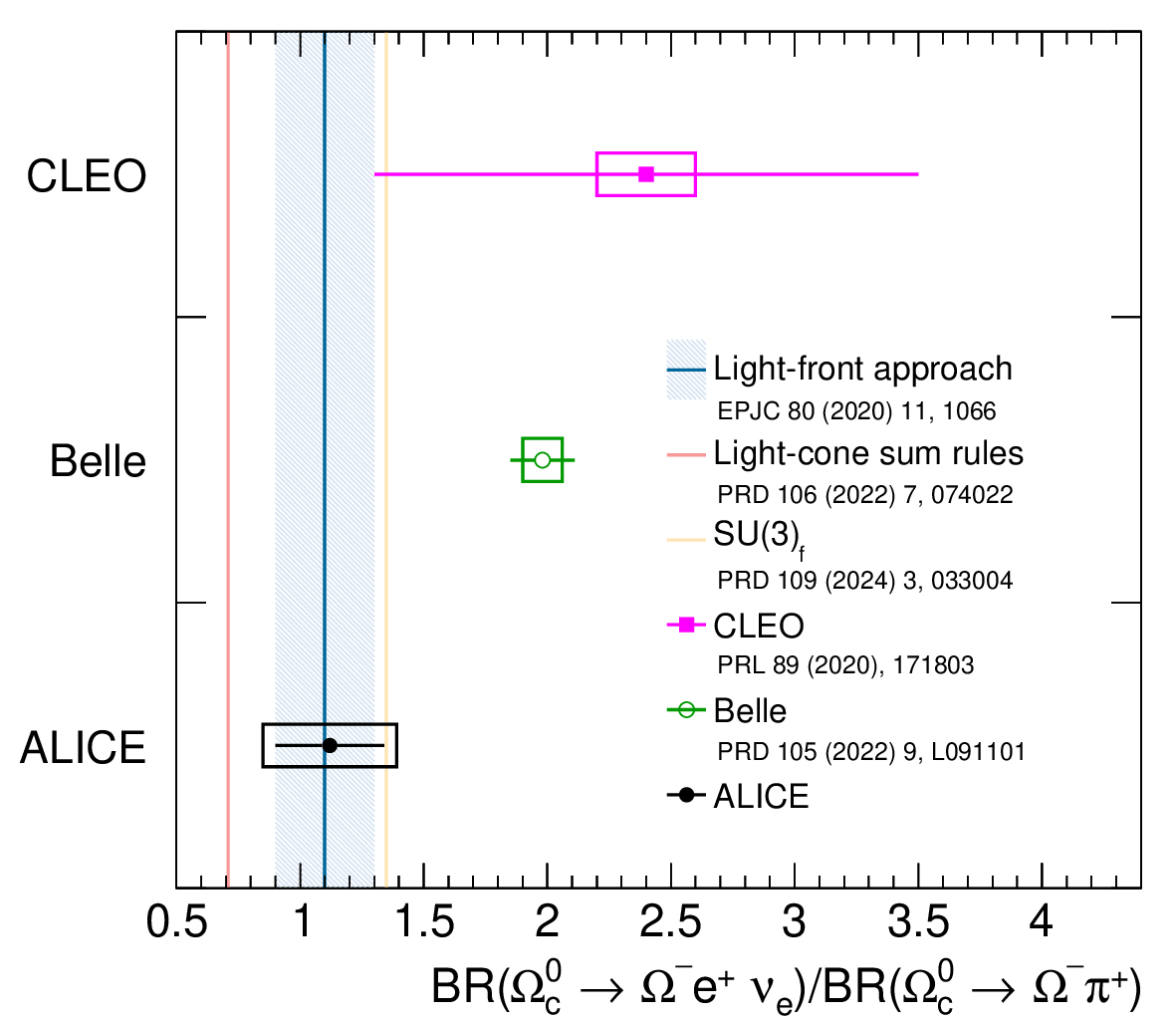}
\caption{Comparison of ${\rm BR}(\Omega^0_{\rm c} \rightarrow \Omega^{-}{\rm e}^{+}\nu_{\rm e})/ {\rm BR}(\Omega^0_{\rm c} \rightarrow \Omega^{-}{\pi}^{+})$ between experiments and theoretical calculations~\cite{CLEO:2002imi, Belle:2021dgc, Hsiao:2020gtc,Aliev:2022gxi,Liu:2023dvg}.}
\label{fig4}
\end{figure}

The ratio of the two measurements, shown in the bottom panel of Fig.~\ref{fig3}, is used to calculate the $p_{\rm T}$ independent branching-fraction ratio. The result was averaged over $p_{\rm T}$ using the inverse uncorrelated relative uncertainties as weights~\cite{weights}. The weights were defined as the sum in quadrature of the relative statistical uncertainties and the $p_{\rm T}$-uncorrelated part of the systematic uncertainties. All the systematic uncertainties were considered as $p_{\rm T}$-correlated in the semileptonic decay. For the hadronic decay, all systematic uncertainties were considered as $p_{\rm T}$-correlated, except for the raw yield extraction. The $p_{\rm T}$-correlated systematic uncertainties were propagated by recomputing the ratio after shifting up and down the ratios with the corresponding $p_{\rm T}$-correlated systematic uncertainties. The final systematic uncertainty on the ratio is obtained by summing the $p_{\rm T}$-correlated and uncorrelated systematic uncertainties in quadrature. The measured ratio is ${\rm BR}(\Omega^0_{\rm c} \rightarrow \Omega^{-}{\rm e}^{+}\nu_{\rm e})/ {\rm BR}(\Omega^0_{\rm c} \rightarrow \Omega^{-}{\pi}^{+})$ = 1.12 $\pm$ 0.22 (stat.) $\pm$ 0.27 (syst.).

In Fig.~\ref{fig4}, the measured \pt-independent branching-fraction ratio is compared with previous experimental measurements from the CLEO Collaboration~\cite{CLEO:2002imi} and Belle Collaboration~\cite{Belle:2021dgc}, and with the theory predictions based on the light-front approach and light-cone sum rules calculations~\cite{Hsiao:2020gtc,Aliev:2022gxi}. The ALICE result is compatible within $1\sigma$ with the CLEO result and is $2.3\sigma$ lower than the one measured by the Belle Collaboration. The ALICE measurement is also consistent within $1\sigma$ with the available theoretical predictions, which showed some tensions with the Belle results.
The present result is also compatible within the uncertainties with the ${\rm BR}(\Xi^0_{\rm c} \rightarrow \Xi^{-}{\rm e}^{+}\nu_{\rm e})/ {\rm BR}(\Xi^0_{\rm c} \rightarrow \Xi^{-}{\pi}^{+})$ measured by the ALICE Collaboration~\cite{ALICE:2021bli}. The agreement between those two measurements is also predicted by the light-front approach calculations~\cite{Hsiao:2020gtc,Zhao:2018zcb}. 
More precise measurements are expected to be performed during Runs 3 and 4 of the LHC. In view of those future measurements, it would be beneficial to compare also with additional model calculations, like LQCD~\cite{Zhang:2021oja} and RQM~\cite{Faustov:2019ddj}, which already provide their prediction for the branching-fraction ratio ${\rm BR}(\Xi^0_{\rm c} \rightarrow \Xi^{-}{\rm e}^{+}\nu_{\rm e})/ {\rm BR}(\Xi^0_{\rm c} \rightarrow \Xi^{-}{\pi}^{+})$.

%%%======================================================================================
\section{Summary}

The inclusive $p_{\rm T}$-differential production cross section of the charm-baryon \Omegac multiplied by the branching ratio
${\rm BR}(\Omega^0_{\rm c} \rightarrow \Omega^{-}{\rm e}^{+}\nu_{\rm e})$ is measured for the first time at midrapidity ($|y| < 0.8$), in the \pt interval $2 < \pt < 12$ GeV$/c$, in pp collisions at \tevt.  
The ${\rm BR}(\Omega^0_{\rm c} \rightarrow \Omega^{-}{\rm e}^{+}\nu_{\rm e})/ {\rm BR}(\Omega^0_{\rm c} \rightarrow \Omega^{-}{\pi}^{+})$ is measured to be 1.12 $\pm$ 0.22 (stat.) $\pm$ 0.27 (syst.), using the inclusive production cross section measured in Ref.~\cite{ALICE:OmegaC}. The branching-fraction ratio is consistent with theory calculations and is 2.3$\sigma$ lower than the value reported by the Belle Collaboration~\cite{Belle:2021dgc}.

%%%%%%%%%%%%%%%%%%%%%%%%%%%%%%%%
% end main text 
%%%%%%%%%%%%%%%%%%%%%%%%%%%%%%%%

%%%%% acknowledgements - handled by EB chairs 
\newenvironment{acknowledgement}{\relax}{\relax}
\begin{acknowledgement}
\section*{Acknowledgements}
% add specific acknowledgements here 
% ...but please don't remove the line below: funding agencies
% will be acknowledged with a custom tex file handled by EB chairs after Collab Round 2
% Version: 2024-04-10

The ALICE Collaboration would like to thank all its engineers and technicians for their invaluable contributions to the construction of the experiment and the CERN accelerator teams for the outstanding performance of the LHC complex.
The ALICE Collaboration gratefully acknowledges the resources and support provided by all Grid centres and the Worldwide LHC Computing Grid (WLCG) collaboration.
The ALICE Collaboration acknowledges the following funding agencies for their support in building and running the ALICE detector:
A. I. Alikhanyan National Science Laboratory (Yerevan Physics Institute) Foundation (ANSL), State Committee of Science and World Federation of Scientists (WFS), Armenia;
Austrian Academy of Sciences, Austrian Science Fund (FWF): [M 2467-N36] and Nationalstiftung f\"{u}r Forschung, Technologie und Entwicklung, Austria;
Ministry of Communications and High Technologies, National Nuclear Research Center, Azerbaijan;
Conselho Nacional de Desenvolvimento Cient\'{\i}fico e Tecnol\'{o}gico (CNPq), Financiadora de Estudos e Projetos (Finep), Funda\c{c}\~{a}o de Amparo \`{a} Pesquisa do Estado de S\~{a}o Paulo (FAPESP) and Universidade Federal do Rio Grande do Sul (UFRGS), Brazil;
Bulgarian Ministry of Education and Science, within the National Roadmap for Research Infrastructures 2020-2027 (object CERN), Bulgaria;
Ministry of Education of China (MOEC) , Ministry of Science \& Technology of China (MSTC) and National Natural Science Foundation of China (NSFC), China;
Ministry of Science and Education and Croatian Science Foundation, Croatia;
Centro de Aplicaciones Tecnol\'{o}gicas y Desarrollo Nuclear (CEADEN), Cubaenerg\'{\i}a, Cuba;
Ministry of Education, Youth and Sports of the Czech Republic, Czech Republic;
The Danish Council for Independent Research | Natural Sciences, the VILLUM FONDEN and Danish National Research Foundation (DNRF), Denmark;
Helsinki Institute of Physics (HIP), Finland;
Commissariat \`{a} l'Energie Atomique (CEA) and Institut National de Physique Nucl\'{e}aire et de Physique des Particules (IN2P3) and Centre National de la Recherche Scientifique (CNRS), France;
Bundesministerium f\"{u}r Bildung und Forschung (BMBF) and GSI Helmholtzzentrum f\"{u}r Schwerionenforschung GmbH, Germany;
General Secretariat for Research and Technology, Ministry of Education, Research and Religions, Greece;
National Research, Development and Innovation Office, Hungary;
Department of Atomic Energy Government of India (DAE), Department of Science and Technology, Government of India (DST), University Grants Commission, Government of India (UGC) and Council of Scientific and Industrial Research (CSIR), India;
National Research and Innovation Agency - BRIN, Indonesia;
Istituto Nazionale di Fisica Nucleare (INFN), Italy;
Japanese Ministry of Education, Culture, Sports, Science and Technology (MEXT) and Japan Society for the Promotion of Science (JSPS) KAKENHI, Japan;
Consejo Nacional de Ciencia (CONACYT) y Tecnolog\'{i}a, through Fondo de Cooperaci\'{o}n Internacional en Ciencia y Tecnolog\'{i}a (FONCICYT) and Direcci\'{o}n General de Asuntos del Personal Academico (DGAPA), Mexico;
Nederlandse Organisatie voor Wetenschappelijk Onderzoek (NWO), Netherlands;
The Research Council of Norway, Norway;
Pontificia Universidad Cat\'{o}lica del Per\'{u}, Peru;
Ministry of Education and Science, National Science Centre and WUT ID-UB, Poland;
Korea Institute of Science and Technology Information and National Research Foundation of Korea (NRF), Republic of Korea;
Ministry of Education and Scientific Research, Institute of Atomic Physics, Ministry of Research and Innovation and Institute of Atomic Physics and Universitatea Nationala de Stiinta si Tehnologie Politehnica Bucuresti, Romania;
Ministry of Education, Science, Research and Sport of the Slovak Republic, Slovakia;
National Research Foundation of South Africa, South Africa;
Swedish Research Council (VR) and Knut \& Alice Wallenberg Foundation (KAW), Sweden;
European Organization for Nuclear Research, Switzerland;
Suranaree University of Technology (SUT), National Science and Technology Development Agency (NSTDA) and National Science, Research and Innovation Fund (NSRF via PMU-B B05F650021), Thailand;
Turkish Energy, Nuclear and Mineral Research Agency (TENMAK), Turkey;
National Academy of  Sciences of Ukraine, Ukraine;
Science and Technology Facilities Council (STFC), United Kingdom;
National Science Foundation of the United States of America (NSF) and United States Department of Energy, Office of Nuclear Physics (DOE NP), United States of America.
In addition, individual groups or members have received support from:
Czech Science Foundation (grant no. 23-07499S), Czech Republic;
European Research Council (grant no. 950692), European Union;
ICSC - Centro Nazionale di Ricerca in High Performance Computing, Big Data and Quantum Computing, European Union - NextGenerationEU;
Academy of Finland (Center of Excellence in Quark Matter) (grant nos. 346327, 346328), Finland.

\end{acknowledgement}

%%%%%%%% Bibliography 
\bibliographystyle{utphys}   % Remember we use title in the biblio
\bibliography{bibliography}
%\input {bibliography.tex}  

%%%%%%%%%%%%%%%%%%%%%%%%%%%%%%%%
% Appendices: yours (if any) + authorlist
%%%%%%%%%%%%%%%%%%%%%%%%%%%%%%%%
\newpage
\appendix

%
%\input{} % put your appendices here (if any)
%

%%%%% Authorlist - please do not touch: handled by EB chairs 
\section{The ALICE Collaboration}
\label{app:collab}
% ALICE Collaboration author list for 2024-04-10
\begin{flushleft} 
\small

S.~Acharya\,\orcidlink{0000-0002-9213-5329}\,$^{\rm 127}$, 
D.~Adamov\'{a}\,\orcidlink{0000-0002-0504-7428}\,$^{\rm 86}$, 
A.~Agarwal$^{\rm 135}$, 
G.~Aglieri Rinella\,\orcidlink{0000-0002-9611-3696}\,$^{\rm 32}$, 
L.~Aglietta$^{\rm 24}$, 
M.~Agnello\,\orcidlink{0000-0002-0760-5075}\,$^{\rm 29}$, 
N.~Agrawal\,\orcidlink{0000-0003-0348-9836}\,$^{\rm 25}$, 
Z.~Ahammed\,\orcidlink{0000-0001-5241-7412}\,$^{\rm 135}$, 
S.~Ahmad\,\orcidlink{0000-0003-0497-5705}\,$^{\rm 15}$, 
S.U.~Ahn\,\orcidlink{0000-0001-8847-489X}\,$^{\rm 71}$
% ...

\section*{Affiliation Notes}

$^{\rm I}$ Deceased\\
$^{\rm II}$ Also at: Max-Planck-Institut fur Physik, Munich, Germany\\
$^{\rm III}$ Also at: Italian National Agency for New Technologies, Energy and Sustainable Economic Development (ENEA), Bologna, Italy\\
$^{\rm IV}$ Also at: Dipartimento DET del Politecnico di Torino, Turin, Italy\\
$^{\rm V}$ Also at: Yildiz Technical University, Istanbul, T\"{u}rkiye\\
$^{\rm VI}$ Also at: Department of Applied Physics, Aligarh Muslim University, Aligarh, India\\
$^{\rm VII}$ Also at: Institute of Theoretical Physics, University of Wroclaw, Poland\\
$^{\rm VIII}$ Also at: An institution covered by a cooperation agreement with CERN\\

\section*{Collaboration Institutes}

$^{1}$ A.I. Alikhanyan National Science Laboratory (Yerevan Physics Institute) Foundation, Yerevan, Armenia\\
$^{2}$ AGH University of Krakow, Cracow, Poland\\
$^{3}$ Bogolyubov Institute for Theoretical Physics, National Academy of Sciences of Ukraine, Kiev, Ukraine\\
$^{4}$ Bose Institute, Department of Physics  and Centre for Astroparticle Physics and Space Science (CAPSS), Kolkata, India\\
$^{5}$ California Polytechnic State University, San Luis Obispo, California, United States\\
$^{6}$ Central China Normal University, Wuhan, China\\
$^{7}$ Centro de Aplicaciones Tecnol\'{o}gicas y Desarrollo Nuclear (CEADEN), Havana, Cuba\\
$^{8}$ Centro de Investigaci\'{o}n y de Estudios Avanzados (CINVESTAV), Mexico City and M\'{e}rida, Mexico\\
$^{9}$ Chicago State University, Chicago, Illinois, United States\\
$^{10}$ China Institute of Atomic Energy, Beijing, China\\
$^{11}$ China University of Geosciences, Wuhan, China\\
$^{12}$ Chungbuk National University, Cheongju, Republic of Korea\\
$^{13}$ Comenius University Bratislava, Faculty of Mathematics, Physics and Informatics, Bratislava, Slovak Republic\\
$^{14}$ Creighton University, Omaha, Nebraska, United States\\
$^{15}$ Department of Physics, Aligarh Muslim University, Aligarh, India\\
$^{16}$ Department of Physics, Pusan National University, Pusan, Republic of Korea\\
$^{17}$ Department of Physics, Sejong University, Seoul, Republic of Korea\\
$^{18}$ Department of Physics, University of California, Berkeley, California, United States\\
$^{19}$ Department of Physics, University of Oslo, Oslo, Norway\\
$^{20}$ Department of Physics and Technology, University of Bergen, Bergen, Norway\\
$^{21}$ Dipartimento di Fisica, Universit\`{a} di Pavia, Pavia, Italy\\
$^{22}$ Dipartimento di Fisica dell'Universit\`{a} and Sezione INFN, Cagliari, Italy\\
$^{23}$ Dipartimento di Fisica dell'Universit\`{a} and Sezione INFN, Trieste, Italy\\
$^{24}$ Dipartimento di Fisica dell'Universit\`{a} and Sezione INFN, Turin, Italy\\
$^{25}$ Dipartimento di Fisica e Astronomia dell'Universit\`{a} and Sezione INFN, Bologna, Italy\\
$^{26}$ Dipartimento di Fisica e Astronomia dell'Universit\`{a} and Sezione INFN, Catania, Italy\\
$^{27}$ Dipartimento di Fisica e Astronomia dell'Universit\`{a} and Sezione INFN, Padova, Italy\\
$^{28}$ Dipartimento di Fisica `E.R.~Caianiello' dell'Universit\`{a} and Gruppo Collegato INFN, Salerno, Italy\\
$^{29}$ Dipartimento DISAT del Politecnico and Sezione INFN, Turin, Italy\\
$^{30}$ Dipartimento di Scienze MIFT, Universit\`{a} di Messina, Messina, Italy\\
$^{31}$ Dipartimento Interateneo di Fisica `M.~Merlin' and Sezione INFN, Bari, Italy\\
$^{32}$ European Organization for Nuclear Research (CERN), Geneva, Switzerland\\
$^{33}$ Faculty of Electrical Engineering, Mechanical Engineering and Naval Architecture, University of Split, Split, Croatia\\
$^{34}$ Faculty of Engineering and Science, Western Norway University of Applied Sciences, Bergen, Norway\\
$^{35}$ Faculty of Nuclear Sciences and Physical Engineering, Czech Technical University in Prague, Prague, Czech Republic\\
$^{36}$ Faculty of Physics, Sofia University, Sofia, Bulgaria\\
$^{37}$ Faculty of Science, P.J.~\v{S}af\'{a}rik University, Ko\v{s}ice, Slovak Republic\\
$^{38}$ Frankfurt Institute for Advanced Studies, Johann Wolfgang Goethe-Universit\"{a}t Frankfurt, Frankfurt, Germany\\
$^{39}$ Fudan University, Shanghai, China\\
$^{40}$ Gangneung-Wonju National University, Gangneung, Republic of Korea\\
$^{41}$ Gauhati University, Department of Physics, Guwahati, India\\
$^{42}$ Helmholtz-Institut f\"{u}r Strahlen- und Kernphysik, Rheinische Friedrich-Wilhelms-Universit\"{a}t Bonn, Bonn, Germany\\
$^{43}$ Helsinki Institute of Physics (HIP), Helsinki, Finland\\
$^{44}$ High Energy Physics Group,  Universidad Aut\'{o}noma de Puebla, Puebla, Mexico\\
$^{45}$ Horia Hulubei National Institute of Physics and Nuclear Engineering, Bucharest, Romania\\
$^{46}$ HUN-REN Wigner Research Centre for Physics, Budapest, Hungary\\
$^{47}$ Indian Institute of Technology Bombay (IIT), Mumbai, India\\
$^{48}$ Indian Institute of Technology Indore, Indore, India\\
$^{49}$ INFN, Laboratori Nazionali di Frascati, Frascati, Italy\\
$^{50}$ INFN, Sezione di Bari, Bari, Italy\\
$^{51}$ INFN, Sezione di Bologna, Bologna, Italy\\
$^{52}$ INFN, Sezione di Cagliari, Cagliari, Italy\\
$^{53}$ INFN, Sezione di Catania, Catania, Italy\\
$^{54}$ INFN, Sezione di Padova, Padova, Italy\\
$^{55}$ INFN, Sezione di Pavia, Pavia, Italy\\
$^{56}$ INFN, Sezione di Torino, Turin, Italy\\
$^{57}$ INFN, Sezione di Trieste, Trieste, Italy\\
$^{58}$ Inha University, Incheon, Republic of Korea\\
$^{59}$ Institute for Gravitational and Subatomic Physics (GRASP), Utrecht University/Nikhef, Utrecht, Netherlands\\
$^{60}$ Institute of Experimental Physics, Slovak Academy of Sciences, Ko\v{s}ice, Slovak Republic\\
$^{61}$ Institute of Physics, Homi Bhabha National Institute, Bhubaneswar, India\\
$^{62}$ Institute of Physics of the Czech Academy of Sciences, Prague, Czech Republic\\
$^{63}$ Institute of Space Science (ISS), Bucharest, Romania\\
$^{64}$ Institut f\"{u}r Kernphysik, Johann Wolfgang Goethe-Universit\"{a}t Frankfurt, Frankfurt, Germany\\
$^{65}$ Instituto de Ciencias Nucleares, Universidad Nacional Aut\'{o}noma de M\'{e}xico, Mexico City, Mexico\\
$^{66}$ Instituto de F\'{i}sica, Universidade Federal do Rio Grande do Sul (UFRGS), Porto Alegre, Brazil\\
$^{67}$ Instituto de F\'{\i}sica, Universidad Nacional Aut\'{o}noma de M\'{e}xico, Mexico City, Mexico\\
$^{68}$ iThemba LABS, National Research Foundation, Somerset West, South Africa\\
$^{69}$ Jeonbuk National University, Jeonju, Republic of Korea\\
$^{70}$ Johann-Wolfgang-Goethe Universit\"{a}t Frankfurt Institut f\"{u}r Informatik, Fachbereich Informatik und Mathematik, Frankfurt, Germany\\
$^{71}$ Korea Institute of Science and Technology Information, Daejeon, Republic of Korea\\
$^{72}$ KTO Karatay University, Konya, Turkey\\
$^{73}$ Laboratoire de Physique Subatomique et de Cosmologie, Universit\'{e} Grenoble-Alpes, CNRS-IN2P3, Grenoble, France\\
$^{74}$ Lawrence Berkeley National Laboratory, Berkeley, California, United States\\
$^{75}$ Lund University Department of Physics, Division of Particle Physics, Lund, Sweden\\
$^{76}$ Nagasaki Institute of Applied Science, Nagasaki, Japan\\
$^{77}$ Nara Women{'}s University (NWU), Nara, Japan\\
$^{78}$ National and Kapodistrian University of Athens, School of Science, Department of Physics , Athens, Greece\\
$^{79}$ National Centre for Nuclear Research, Warsaw, Poland\\
$^{80}$ National Institute of Science Education and Research, Homi Bhabha National Institute, Jatni, India\\
$^{81}$ National Nuclear Research Center, Baku, Azerbaijan\\
$^{82}$ National Research and Innovation Agency - BRIN, Jakarta, Indonesia\\
$^{83}$ Niels Bohr Institute, University of Copenhagen, Copenhagen, Denmark\\
$^{84}$ Nikhef, National institute for subatomic physics, Amsterdam, Netherlands\\
$^{85}$ Nuclear Physics Group, STFC Daresbury Laboratory, Daresbury, United Kingdom\\
$^{86}$ Nuclear Physics Institute of the Czech Academy of Sciences, Husinec-\v{R}e\v{z}, Czech Republic\\
$^{87}$ Oak Ridge National Laboratory, Oak Ridge, Tennessee, United States\\
$^{88}$ Ohio State University, Columbus, Ohio, United States\\
$^{89}$ Physics department, Faculty of science, University of Zagreb, Zagreb, Croatia\\
$^{90}$ Physics Department, Panjab University, Chandigarh, India\\
$^{91}$ Physics Department, University of Jammu, Jammu, India\\
$^{92}$ Physics Program and International Institute for Sustainability with Knotted Chiral Meta Matter (SKCM2), Hiroshima University, Hiroshima, Japan\\
$^{93}$ Physikalisches Institut, Eberhard-Karls-Universit\"{a}t T\"{u}bingen, T\"{u}bingen, Germany\\
$^{94}$ Physikalisches Institut, Ruprecht-Karls-Universit\"{a}t Heidelberg, Heidelberg, Germany\\
$^{95}$ Physik Department, Technische Universit\"{a}t M\"{u}nchen, Munich, Germany\\
$^{96}$ Politecnico di Bari and Sezione INFN, Bari, Italy\\
$^{97}$ Research Division and ExtreMe Matter Institute EMMI, GSI Helmholtzzentrum f\"ur Schwerionenforschung GmbH, Darmstadt, Germany\\
$^{98}$ Saga University, Saga, Japan\\
$^{99}$ Saha Institute of Nuclear Physics, Homi Bhabha National Institute, Kolkata, India\\
$^{100}$ School of Physics and Astronomy, University of Birmingham, Birmingham, United Kingdom\\
$^{101}$ Secci\'{o}n F\'{\i}sica, Departamento de Ciencias, Pontificia Universidad Cat\'{o}lica del Per\'{u}, Lima, Peru\\
$^{102}$ Stefan Meyer Institut f\"{u}r Subatomare Physik (SMI), Vienna, Austria\\
$^{103}$ SUBATECH, IMT Atlantique, Nantes Universit\'{e}, CNRS-IN2P3, Nantes, France\\
$^{104}$ Sungkyunkwan University, Suwon City, Republic of Korea\\
$^{105}$ Suranaree University of Technology, Nakhon Ratchasima, Thailand\\
$^{106}$ Technical University of Ko\v{s}ice, Ko\v{s}ice, Slovak Republic\\
$^{107}$ The Henryk Niewodniczanski Institute of Nuclear Physics, Polish Academy of Sciences, Cracow, Poland\\
$^{108}$ The University of Texas at Austin, Austin, Texas, United States\\
$^{109}$ Universidad Aut\'{o}noma de Sinaloa, Culiac\'{a}n, Mexico\\
$^{110}$ Universidade de S\~{a}o Paulo (USP), S\~{a}o Paulo, Brazil\\
$^{111}$ Universidade Estadual de Campinas (UNICAMP), Campinas, Brazil\\
$^{112}$ Universidade Federal do ABC, Santo Andre, Brazil\\
$^{113}$ Universitatea Nationala de Stiinta si Tehnologie Politehnica Bucuresti, Bucharest, Romania\\
$^{114}$ University of Cape Town, Cape Town, South Africa\\
$^{115}$ University of Derby, Derby, United Kingdom\\
$^{116}$ University of Houston, Houston, Texas, United States\\
$^{117}$ University of Jyv\"{a}skyl\"{a}, Jyv\"{a}skyl\"{a}, Finland\\
$^{118}$ University of Kansas, Lawrence, Kansas, United States\\
$^{119}$ University of Liverpool, Liverpool, United Kingdom\\
$^{120}$ University of Science and Technology of China, Hefei, China\\
$^{121}$ University of South-Eastern Norway, Kongsberg, Norway\\
$^{122}$ University of Tennessee, Knoxville, Tennessee, United States\\
$^{123}$ University of the Witwatersrand, Johannesburg, South Africa\\
$^{124}$ University of Tokyo, Tokyo, Japan\\
$^{125}$ University of Tsukuba, Tsukuba, Japan\\
$^{126}$ Universit\"{a}t M\"{u}nster, Institut f\"{u}r Kernphysik, M\"{u}nster, Germany\\
$^{127}$ Universit\'{e} Clermont Auvergne, CNRS/IN2P3, LPC, Clermont-Ferrand, France\\
$^{128}$ Universit\'{e} de Lyon, CNRS/IN2P3, Institut de Physique des 2 Infinis de Lyon, Lyon, France\\
$^{129}$ Universit\'{e} de Strasbourg, CNRS, IPHC UMR 7178, F-67000 Strasbourg, France, Strasbourg, France\\
$^{130}$ Universit\'{e} Paris-Saclay, Centre d'Etudes de Saclay (CEA), IRFU, D\'{e}partment de Physique Nucl\'{e}aire (DPhN), Saclay, France\\
$^{131}$ Universit\'{e}  Paris-Saclay, CNRS/IN2P3, IJCLab, Orsay, France\\
$^{132}$ Universit\`{a} degli Studi di Foggia, Foggia, Italy\\
$^{133}$ Universit\`{a} del Piemonte Orientale, Vercelli, Italy\\
$^{134}$ Universit\`{a} di Brescia, Brescia, Italy\\
$^{135}$ Variable Energy Cyclotron Centre, Homi Bhabha National Institute, Kolkata, India\\
$^{136}$ Warsaw University of Technology, Warsaw, Poland\\
$^{137}$ Wayne State University, Detroit, Michigan, United States\\
$^{138}$ Yale University, New Haven, Connecticut, United States\\
$^{139}$ Yonsei University, Seoul, Republic of Korea\\
$^{140}$  Zentrum  f\"{u}r Technologie und Transfer (ZTT), Worms, Germany\\
$^{141}$ Affiliated with an institute covered by a cooperation agreement with CERN\\
$^{142}$ Affiliated with an international laboratory covered by a cooperation agreement with CERN.\\

\end{flushleft} 
  
\end{document}